\documentclass[prl,aps,twocolumn,superscriptaddress,lengthcheck,showpacs,amsmath,amssymb,floatfix]{revtex4-1}

\usepackage{microtype}
\usepackage{graphicx}
\usepackage[rgb]{xcolor}
\usepackage{hyperref}
\hypersetup{colorlinks,linkcolor=blue,citecolor=blue,urlcolor=blue}

\newcommand{\bra}[1]{\langle #1 |}
\newcommand{\ket}[1]{| #1 \rangle}

\DeclareMathOperator*{\SumInt}{%
	\mathchoice%
	{\ooalign{$\displaystyle\sum$\cr\hidewidth$\displaystyle\int$\hidewidth\cr}}
	{\ooalign{\raisebox{.14\height}{\scalebox{.7}{$\textstyle\sum$}}\cr\hidewidth$\textstyle\int$\hidewidth\cr}}
	{\ooalign{\raisebox{.2\height}{\scalebox{.6}{$\scriptstyle\sum$}}\cr$\scriptstyle\int$\cr}}
	{\ooalign{\raisebox{.2\height}{\scalebox{.6}{$\scriptstyle\sum$}}\cr$\scriptstyle\int$\cr}}
}

\newcommand{\appropto}{\mathrel{\vcenter{
			\offinterlineskip\halign{\hfil$##$\cr
				\propto\cr\noalign{\kern2pt}\sim\cr\noalign{\kern-2pt}}}}}

\begin{document}

\title{Signatures of Vibrational Strong Coupling in Raman Scattering}
\author{Javier del Pino}
\affiliation{Departamento de F{\'\i}sica Te{\'o}rica de la Materia Condensada and Condensed Matter Physics Center (IFIMAC), Universidad Aut\'onoma de Madrid, Madrid E-28049, Spain}
\author{Johannes Feist}
\affiliation{Departamento de F{\'\i}sica Te{\'o}rica de la Materia Condensada and Condensed Matter Physics Center (IFIMAC), Universidad Aut\'onoma de Madrid, Madrid E-28049, Spain}
\author{F.~J.~Garcia-Vidal}
\affiliation{Departamento de F{\'\i}sica Te{\'o}rica de la Materia Condensada and Condensed Matter Physics Center (IFIMAC), Universidad Aut\'onoma de Madrid, Madrid E-28049, Spain}
\affiliation{Donostia International Physics Center (DIPC), E-20018 Donostia/San Sebastian, Spain}

\begin{abstract}
In this Letter we analyze theoretically how the emergence of collective strong coupling between vibrational excitations and confined cavity modes affects Raman scattering processes. This work is motivated by recent 
experiments [Shalabney \emph{et al.}, Angew. Chemie 54, 7971 (2015)], which reported enhancements of up to three orders of magnitude in the Raman signal. By using different models within linear response theory, 
we show that the total Raman cross section is maintained constant when the system evolves from the weak-coupling limit to the strong-coupling regime. A redistribution of the Raman signal among the two polaritons 
is the main fingerprint of vibrational strong coupling in the Raman spectrum. 
\end{abstract}


\maketitle

Raman scattering is one of the principal methods used to obtain information about material properties and chemical structure~\cite{Colthup1990}. In particular, it probes the rovibrational structure of matter and can thus be used to provide a ``fingerprint'', making it useful for a wide range of applications in research and industry. Its operating principle relies on the inelastic scattering of optical photons (frequency $\omega_L$), which leads to the emission of photons at shifted frequencies $\omega_L - \delta\omega$. The observed frequency shifts $\delta\omega$ correspond to Raman-allowed excitations in the system under study and thus provide detailed information about its (rovibrational) states. Vibrational excitations are often well-approximated by harmonic oscillators, leading to a series of equidistant \emph{Stokes} lines $\delta\omega = n\omega_v$ for each mode, corresponding to excitation of $n$ vibrational quanta.

Recently, it has been shown that vibrational excitations interacting with confined cavity modes can enter the vibrational-strong-coupling (VSC) regime~\cite{Shalabney2015,Long2015,George2015,DelPino2015,Simpkins2015}. Strong coupling, already well-known in the context of electronic excitations (see \cite{Carusotto2013,Torma2015} for recent reviews), occurs when the coherent energy exchange between a light mode and matter excitations is faster than the decay and/or decoherence of either constituent. The fundamental excitations of both systems then become inextricably linked and can be described as hybrid light-matter quasiparticles, so-called polaritons, that combine the properties of both constituents. Consequently, the vibro-polaritons obtained under VSC are formed by superpositions of a cavity photon and excited molecular bond vibrations that are collectively distributed over a large number of molecules.

A recent pioneering experiment~\cite{Shalabney2015a} measured spontaneous Raman scattering under collective strong coupling of the (IR- and Raman-active) C=O bond of a polymer (polyvinyl acetate, PVAc) to Fabry-Perot cavity photons. A large increase of the Raman signal under strong coupling was observed, with emission at energy shifts $\delta\omega$ approximately corresponding to the upper and lower polaritons. This intriguing result motivates our current study. We theoretically investigate the signatures of vibrational strong coupling in the Raman spectrum. We mention here that in this phenomenon the \emph{vibrational excitation} is modified by interaction with the cavity, in contrast to the well-known techniques of surface-enhanced Raman scattering~\cite{CamKam1998,KneKneKne2008} that are based on the enhancement of the \emph{optical transitions} through modification of the electromagnetic density of states. As opposed to surface-enhanced Raman scattering, there is no known simple picture that explains a possible enhancement of the Raman cross section under vibrational strong coupling.

\begin{figure}[tb]
\includegraphics[width=0.75\linewidth]{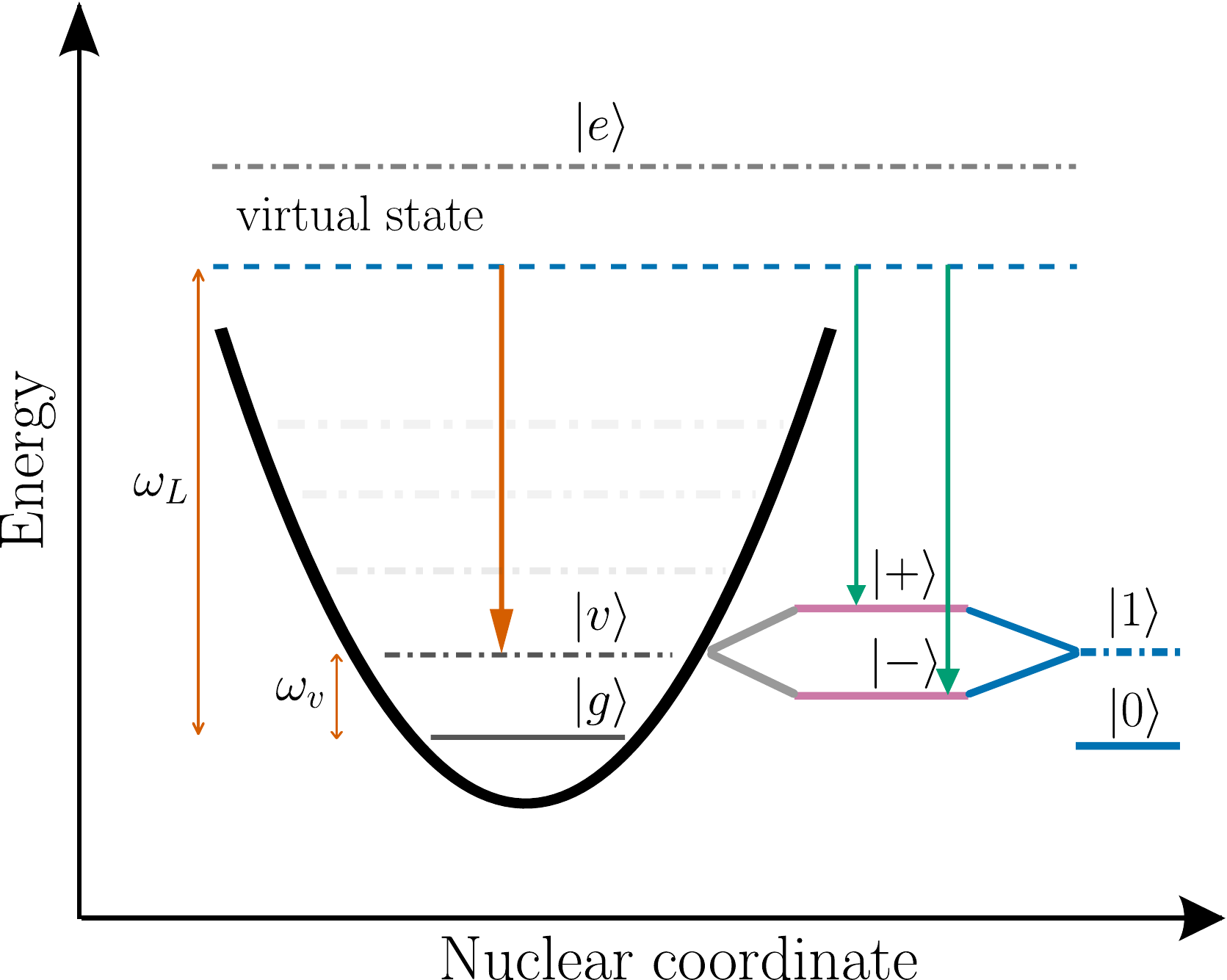}
\caption{Sketch of the Raman scattering process for a molecule as a result of excitation with an off-resonant driving field. After coherent excitation, the molecule promotes to a virtual state depicted with a blue dashed line. In the weak coupling regime, the electron decays into the first excited vibrational state, $\ket{v}$ (orange arrow). When this first excited vibrational state is strongly coupled to the cavity mode, $\ket{1}$, the electron can decay to any of the two polaritons, $\ket{+}$ or $\ket{-}$ (blue arrows).}\label{Fig:1}
\end{figure}	

\autoref{Fig:1} illustrates the Raman scattering process associated with the first Stokes line of a single molecular vibration, in both the weak- and strong-coupling regimes. In our first approach, we restrict the description of the bare molecule to a $\Lambda$-system that comprises the three states involved in this Raman process: the ground state, $\ket{g}$, with energy $\omega_g$, the first excited vibrational mode, $\ket{v}$, with energy $\omega_v$, and the electronically excited state, $\ket{e}$, with energy $\omega_e$. Regarding the description of the cavity field, we only consider the first cavity mode, $\ket{1}$, with energy $\omega_c$. In all the calculations presented in this work, we will consider the case of zero detuning, i.e., $\omega_c=\omega_v$. Choosing $\omega_g$ as the zero energy, the coherent dynamics of the system composed of $N$ molecules and a cavity mode is governed by the Hamiltonian ($\hbar=1$):
\begin{multline}
\hat{H} = \omega_c\ket{1}\bra{1}+ \sum_{i=1}^N [\omega_v \ket{v_i}\bra{v_i} + \omega_e\ket{e_i}\bra{e_i} + \\ g\left(\ket{1}\bra{v_i} + \ket{v_i}\bra{1} \right)],
\end{multline}
where the cavity-oscillator interaction is measured by $g$, which depends on the cavity electric field strength and the change of the molecular dipole moment under displacement from the equilibrium position, but in our calculations will be used as a parameter that fixes the Rabi splitting (see below). For simplicity, we assume a configuration with zero disorder in which all the $N$ molecules are equally coupled to the cavity mode. We have checked that this does not affect the conclusions presented here.

If the probe field is far off-resonant, $\omega_v\ll\omega_L\ll\omega_e$, we can safely work within second-order perturbation theory and neglect losses in the system. In general, the Raman scattering cross section associated with a process where the system is excited from the initial state $\ket{i}$ (energy $\omega_i$) to a final state $\ket{f}$ (energy $\omega_f$, with scattered photon energy $\omega_L - \omega_f + \omega_i$) can be written as $\sigma_{R,\omega_f-\omega_i} \propto \left|\alpha_{fi}\right|^2$ where $\alpha_{f i} = \bra{f} \hat{\alpha} \ket{i} = \bra{f} \hat{\mu} \frac1{\hat{H}-\omega_L-\omega_i} \hat{\mu} \ket{i}$ is the polarizability matrix element between the initial and final states, with $\hat{\mu}$ being the dipole operator and $\hat{H}$ the Hamiltonian. In our case, we only consider dipolar transitions $\ket{g}\rightarrow\ket{e}$, characterized by a dipole moment $\mu_{ge}$, and $\ket{e}\rightarrow\ket{v}$ with dipole moment  $\mu_{ev}$, corresponding to an electronic excitation from the ground state and transition from the electronically excited state to the first excited vibrational mode. For vibrational modes that are IR-active (as required for VSC), there are also direct dipole transitions from the ground state to the vibrationally excited state. However, these do not play a role in Raman scattering to the vibrationally excited modes, which requires two dipole transitions.

In the weak-coupling regime ($g \rightarrow 0$), the Raman scattering process corresponds to an excitation from the global ground state, $\ket{G} = \Pi_{i=1}^N \ket{g_i}$, followed by decay into a singly excited vibrational state of a molecule, $\ket{v_i}$ (shorthand for $\ket{v_i}\prod_{j\neq i}\ket{g_j}$), with index $i=1,\dots, N$ labeling different molecules.  In this situation the molecules act independently and the cross section for emission of a photon of energy $\omega_L-\omega_v$ is just the sum of the cross sections associated with each molecule:
\begin{equation}\label{eq:sigmaraman_WC}
\sigma_{\mathrm{R},\omega_v} \propto \sum_{i=1}^N|\alpha_{v_i G}|^2 = N \left[\frac{\mu_{ve}\mu_{ge}}{\omega_e-\omega_L}\right]^2.
\end{equation}

In the VSC regime, the ($N + 1$) singly excited eigenstates of the system are formed by: \emph{(i)} two \emph{polaritons}, $\ket{\pm}=\frac1{\sqrt{2}}(\ket{1}\pm\ket{B})$, symmetric and antisymmetric linear combinations of the cavity mode, $\ket{1}$, with the collective \emph{bright state} of the molecular excitation, $\ket{B}=\frac1{\sqrt{N}}\sum_{i=1}^N \ket{v_i}$; and \emph{(ii)} the so-called \emph{dark states}, $\ket{d}$, ($N-1$) combinations of molecular excitations orthogonal to $\ket{B}$, which have eigenfrequencies $\omega_v$ and no electromagnetic component. The eigenfrequencies of the two polariton modes are $\omega_{\pm}=\omega_v\pm g\sqrt{N}$, the Rabi splitting being $\Omega_R=2g\sqrt{N}$. In principle, it could be expected that the formation of collective modes among the molecular bonds leads to an enhancement of the Raman cross section. However, a straightforward calculation shows that the Raman cross sections associated with the dark modes are zero whereas those of the two polaritons (involving photons of energies $\omega_L-\omega_+$ and $\omega_L-\omega_-$) are just half of the Raman cross section evaluated in the weak-coupling limit, \autoref{eq:sigmaraman_WC}. In other words, when going from the weak to the strong coupling regime, the total Raman cross section is maintained but equally shared between the two polaritons.

This is an interesting result as it points to the existence of a kind of sum rule for the Raman scattering cross section. In order to investigate this in more detail, we analyze the total Raman scattering cross section defined as the sum over all possible final states, $\ket{f}$, resulting from inelastic processes when the system is excited from the ground state $\ket{G}$:
\begin{equation}\label{eq:totraman_hamilt}
\Sigma_R \propto \SumInt_{f \neq G} \left| \bra{f} \hat{\alpha} \ket{G} \right|^2 = \bra{G} \hat{\alpha}^2 \ket{G} - \bra{G}\hat{\alpha}\ket{G}^2,
\end{equation}
where we have used the closure relation $\SumInt_f \ket{f}\bra{f}=\hat{I}$, which implies $\SumInt_{f \neq G}=\hat{I}-\ket{G}\bra{G}$.
We are interested in the change of the cross section as the Hamiltonian is changed and VSC is established. By inserting the spectral decomposition of the Hamiltonian in $\hat{\alpha} = \hat{\mu} \frac1{\hat{H}-\omega_L-\omega_i} \hat{\mu}$, it can be seen that the total Raman cross section will only be affected by changes in the ground state $\ket{G}$ or intermediate electronically excited states $\ket{n}$ reachable by a single-photon transition, $\bra{n}\hat{\mu}\ket{G}\not=0$. Vibrational strong coupling primarily affects the \emph{final} states, i.e., the vibrationally excited states that split into polaritons. Furthermore, as the driving frequency $\omega_L$ in standard Raman scattering experiments is not close to any eigenstate of the Hamiltonian, any possible changes in the electronically excited states are not expected to have a big effect. This suggests that, within the theoretical framework described above, changes in the total Raman cross section when going from the weak coupling limit to the strong coupling regime could only come from changes in the ground state of the system. Such changes are well-known to occur when the system reaches \emph{ultrastrong} coupling, i.e., when the Rabi splitting $\Omega_R$ becomes comparable to the transition energy $\omega_v$ and counterrotating terms in the emitter-cavity coupling cease to be negligible. It has recently been shown that changes in the ground state properties under ultrastrong coupling depend sensitively on the observable that is interrogated~\cite{Galego2015,Cwik2015}, with e.g., bond-length changes only being sensitive to the single-molecule coupling strength, while energy shifts depend on the collective coupling.

In order to explicitly check the formalism above, as well as go beyond it, we therefore turn to a microscopic quantum model for the organic molecules interacting with the quantized cavity field. In this formalism, we include counterrotating terms to explore the effects of ultrastrong coupling, and additionally incorporate losses and dephasing mechanisms that were not present in the previous approach. The Hamiltonian describing the $i^{\mathrm{th}}$ bare molecule now reads:
\begin{equation}
\hat{H}_{\mathrm{mol}}^{(i)} = \omega_e\hat{\sigma}_i^{\dagger}\hat{\sigma}_i + \omega_v [\hat{b}_i^{\dagger}\hat{b}_i + \sqrt{S}\hat{\sigma}_i^{\dagger}\hat{\sigma}_i(\hat{b}_i^{\dagger} + \hat{b}_i)],
\end{equation}
where the electronic transition of the molecule, of energy $\omega_e$, is described by the Pauli ladder operator $\hat{\sigma}_i$, whereas $\hat{b}_i$ is the annihilation operator of the optically active vibrational mode of the molecule of energy $\omega_v$. The interaction between electronic and vibrational states is characterized by the Huang-Rhys parameter $S$, which quantifies the phonon displacement between the ground and excited electronic states. The total Hamiltonian also contains the cavity field and the coupling term between the cavity mode and the vibrational states of the molecules:
\begin{equation}\label{eq:microscopic_hamiltonian}
\hat{H}=\sum_{i=1}^N \hat{H}_{\mathrm{mol}}^{(i)} + \omega_c \hat{a}^{\dagger}\hat{a} + g\sum_{i=1}^N(\hat{a}+\hat{a}^{\dagger})(\hat{b}_i + \hat{b}_i^{\dagger}),
\end{equation}
where $\hat{a}$ is the annihilation operator for the cavity mode with energy $\omega_c$. As in our previous approach, $g$ describes the cavity-vibrational mode interaction, but notice that in \autoref{eq:microscopic_hamiltonian} we now include counterrotating terms that do not conserve the number of excitations.

To account for both loss and dephasing mechanisms, we rely on the standard Lindblad master equation formalism~\cite{Petruccione2007}. We include decay of the electronic excitations (rate $\gamma_e$) and vibrational modes (rate $\gamma_v$), as well as the loss of the cavity photons (rate $\kappa$). Additionally, we consider elastic scattering with bath modes, which leads to pure electronic (rate $\gamma^{\phi}_{e}$) and vibrational (rate $\gamma^{\phi}_{v}$) dephasing terms. The time evolution of the density matrix $\hat{\rho}$ is then described by
\begin{multline}\label{eq:densmat_prop}
\partial_t \hat{\rho} = -i[\hat{H},\hat{\rho}] + \kappa \mathcal{L}_{\hat{a}} [\hat{\rho}] + \sum_{i=1}^N(\gamma_e\mathcal{L}_{\hat{\sigma}_i} [\hat{\rho}]\\
+ \gamma_v\mathcal{L}_{\hat{b}_i} [\hat{\rho}] +\gamma^{\phi}_{e} \mathcal{L}_{\hat{\sigma}_i^{\dagger}\hat{\sigma}_i}[\hat{\rho}] + \gamma^{\phi}_{v} \mathcal{L}_{\hat{b}_i^\dagger \hat{b}_i}[\hat{\rho}]),
\end{multline}	
where $\mathcal{L}_{\hat{X}} [\hat{\rho}] = \hat{X}\hat{\rho} \hat{X}^{\dagger}  - \frac{1}{2}\{\hat{X}^{\dagger} \hat{X}, \hat{\rho}\}$. We note that in the ultrastrong-coupling regime, the Hamiltonian does not conserve the number of excitations and the terms $\mathcal{L}_{\hat{a}}[\hat{\rho}]$ and $\mathcal{L}_{\hat{b}_i} [\hat{\rho}]$ actually introduce artificial pumping. We then replace these terms by explicitly calculating the decay introduced by coupling to a zero-temperature bath of background modes with constant spectral density, within Bloch-Redfield-Wangsness (BRW) theory~\cite{Wangsness1953,Redfield1955}. As we have previously shown~\cite{DelPino2015}, BRW theory should in principle also be used for the description of vibrational dephasing; however, this does not influence the results presented here significantly, and we thus use Lindblad terms for simplicity.

We additionally introduce an off-resonant laser field at frequency $\omega_L$ that collectively drives all emitters, represented by $\hat{H}_{d}=\Omega_p \sum_i (\hat{\sigma}_ie^{-i\omega_Lt} + \hat{\sigma}_i^{\dagger}e^{i\omega_L t})$. The emission spectrum is then calculated from the steady-state two-time correlation function of the electronic dipole within a frame rotating at $\omega_L$ in which $\hat{H}_d$ is time-independent. This gives the emission spectrum $S(\omega) = \int_{-\infty}^{\infty} e^{i(\omega_L-\omega)\tau} \langle \hat{\sigma}^{\dagger}(\tau) \hat{\sigma}(0) \rangle \mathrm{d}\tau$, where $\hat{\sigma}=\sum_i\hat{\sigma}_i$, from which we remove the zero-frequency Rayleigh peak to obtain only the Raman contribution. For the numerical implementation of this microscopic model, we employ the open-source QuTiP package~\cite{Johansson2013}.

\begin{figure}[tb]
\includegraphics[width=\linewidth]{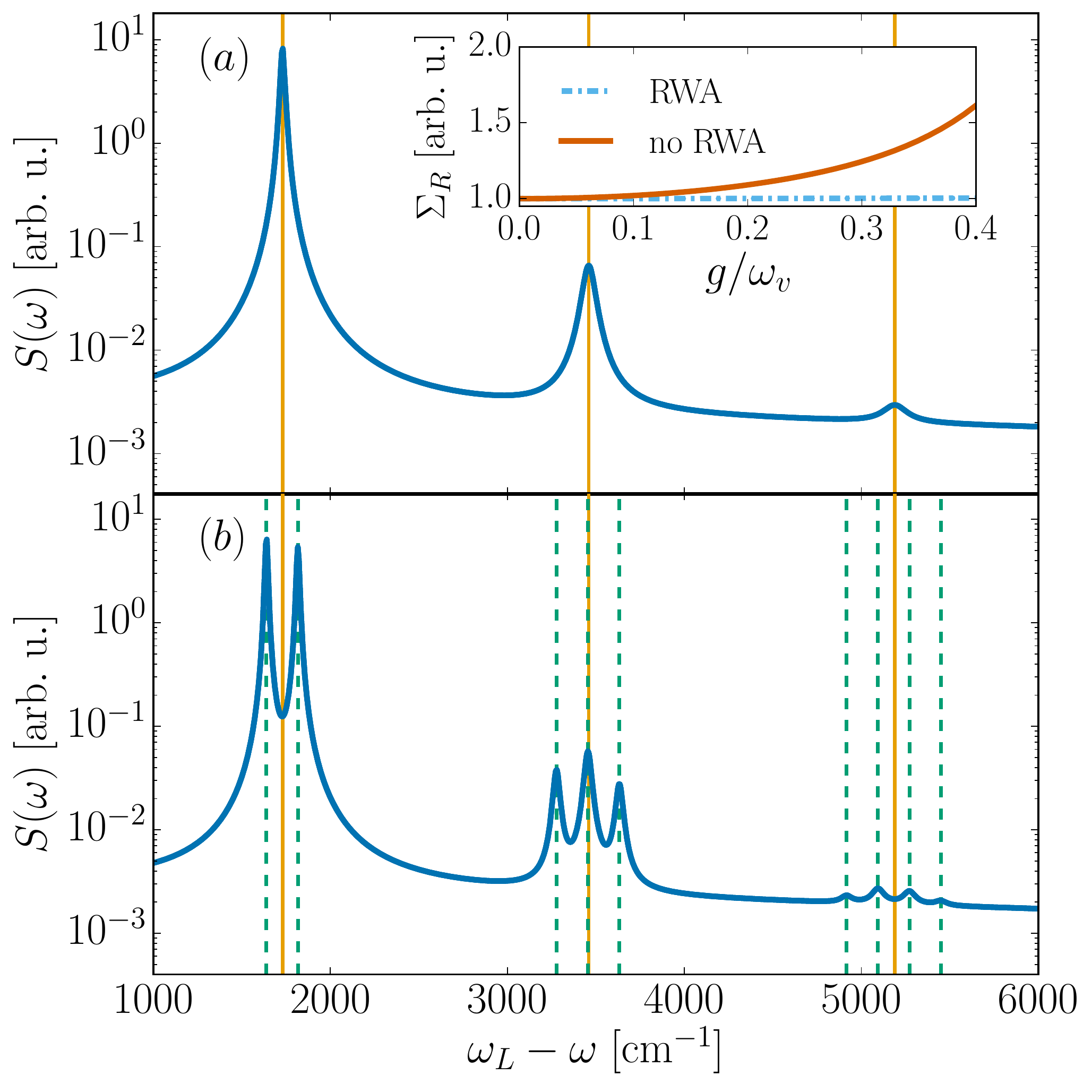}
\caption{Single molecule Raman spectrum in the weak ($a$) and strong coupling ($b$) regimes.  A weak probe strength $\Omega_p\ll\omega_v$ is used in both cases. The Stokes lines at energies  $n\omega_v, n\in \mathbb{N}$ are depicted in orange, while the dashed green lines spectrally located at $\omega^{(n, m)}$ (see main text) show the dressed energies in the strong coupling regime. In this last calculation we allowed for five excitations at most. Inset: Total Raman scattering probability $\int\! S(\omega) \mathrm{d}\omega \propto \Sigma_R$ as a function of cavity-oscillator interaction $g$, within (blue) and without (orange) the rotating wave approximation.}\label{Fig:2}
\end{figure}

We first apply this theoretical framework to study the Raman spectrum when the vibrational mode of a single molecule is coupled to the cavity mode. The results are depicted in \autoref{Fig:2}, with parameters chosen to agree with the experiment~\cite{Shalabney2015a}. The vibrational frequency is $\omega_v = 1730~$cm$^{-1}$ and the vibrational rates, $\gamma_{v} = \gamma_{v}^{\phi}=13~$cm$^{-1}$, are recovered from the experimental transmission spectrum assuming that half of the total linewidth is due to pure dephasing. The cavity losses are accounted for by $\kappa=13~$cm$^{-1}$, which is a relatively small value chosen to make the separate peaks clearly visible. Other parameters are also chosen in accordance with typical values for polymers: $\omega_e=5~$eV, $\omega_L \gg\omega_v$, $S=2$, $\gamma_{e} =50~$cm$^{-1}$ and $\gamma_{e}^{\phi} =50~$cm$^{-1}$. As expected, in the weak-coupling limit, i.e., $g \to 0$ (\autoref{Fig:2}a), Stokes lines appear at the vibrational frequencies, $n \omega_v$, with $n\in\mathbb{N}$. When VSC emerges ($\Omega_R=160~$cm$^{-1}$ in \autoref{Fig:2}b), the Stokes lines split into several sidebands given by $\omega^{(n, m)}=n\omega_- + m\omega_+$ with $n,m\in\mathbb{N}$, where $\omega_\pm = \omega_v \sqrt{1\pm2g/\omega_v}$ when counterrotating terms are included. Importantly, the position of the first Stokes lines coincides with their position in the transmission spectrum, and consequently the splitting between the two peaks coincides with $\Omega_R$, the Rabi splitting that can be measured in the transmission spectrum. In addition, the total Raman signal $\int\!S(\omega)\mathrm{d}\omega\propto\Sigma_R$ stays almost constant when going from the weak to the strong-coupling regime. This result is compatible with the prediction obtained from the pure Hamiltonian approach (\autoref{eq:totraman_hamilt}), as in this case, $\Omega_R \ll \omega_v$.

In order to observe changes in the total Raman signal, the system should enter into the ultrastrong coupling regime. This is shown in the inset of  \autoref{Fig:2}a), in which we render the evolution of the total Raman scattering probability as a function of $g$, with all the other parameters being the same as those used in the main panels of \autoref{Fig:2}. These results demonstrate that under \emph{ultrastrong} coupling, the total Raman cross section could indeed change significantly, as it increases by a factor of around $1.5$ in the limit $g\to\omega_v/2$ (larger values of $g$ are unphysical within this model). Notice that, however, in the experiments~\cite{Shalabney2015a} $g/ \omega_v \approx 0.05$ and our results show that the total Raman scattering probability is practically the same as that obtained in the weak coupling limit, $g \to 0$. 

\begin{figure}[tb]
\includegraphics[width=\linewidth]{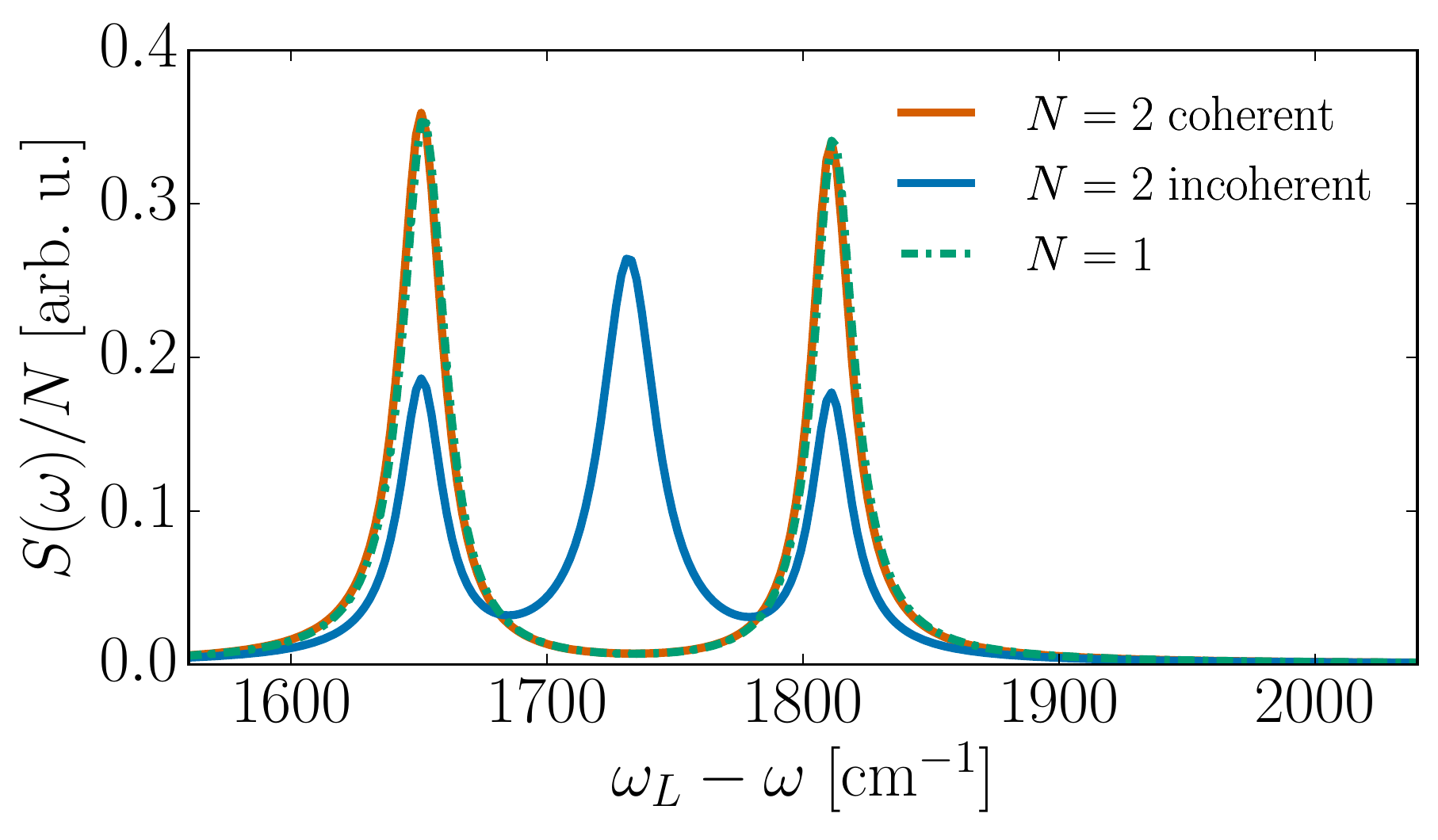}
\caption{First Stokes line of the Raman spectra in the strong coupling regime, for one and two molecules. The chosen parameters are the same as those used for the single-molecule case, with the spectra normalized to the number of molecules $N$. For the case of two molecules, the spectra for coherent and incoherent collection are depicted. For comparison, the case of a single molecule already shown in \autoref{Fig:2}b is also rendered.}\label{Fig:3}
\end{figure}

Our previous results using the microscopic model were obtained for a single molecule. As a minimal model to investigate collective effects, we now show the Raman spectrum of two molecules strongly coupled to a cavity mode. For comparison with the single-molecule case, we rescale $g\to g/\sqrt{2}$ to keep the Rabi frequency $\Omega_R$ constant. The results (see \autoref{Fig:3}) are now also sensitive to the collection operator as different physics arise if we examine either the \emph{coherent} case (obtained from the correlation function of the total dipole operator $\sum_{i=1}^N\hat{\sigma}_i$) or the \emph{incoherent} sum over different molecules $\sum_i S_i(\omega)$ where $S_i(\omega)\propto \int_{-\infty}^{\infty}e^{i(\omega_L-\omega)\tau}\langle \hat{\sigma}_i^{\dagger}(\tau) \hat{\sigma}_i(0) \rangle$ is the Raman spectrum associated with one molecule. For coherent collection of the emission from both molecules, the same spectral weight redistribution among the two polaritons (split by $\Omega_R$, as in transmission measurements) is exhibited. By comparing it with the case of a single molecule (also depicted in the figure) we infer a linear scaling $\propto N$, which is consistent with the $\Lambda$-system results (\autoref{eq:sigmaraman_WC}). This confirms that no collective enhancement of the Raman signal is present. Interestingly, while in coherent collection only the polaritons are observed in the spectrum, for incoherent collection (which could be achieved experimentally using, e.g., a near-field probe), a central peak appears at the bare vibrational energy $\omega_v$, a signature of the vibrational dark state $\ket{d}=\frac1{\sqrt{2}}(b_1^{\dagger}-b_2^{\dagger})\ket{G}$. This demonstrates that the dark state emission is suppressed under coherent collection due to destructive interference (as observed within the $\Lambda$-system approach above), even though these states emit on the single-molecule level. Nonetheless, the total Raman signal $\int\!S(\omega)\mathrm{d}\omega$ is almost independent of the collection method. Combined with the results above, we can thus conclude that under strong coupling, the total dipole strength ($\propto N$) is redistributed between the modes, but not enhanced significantly.

In conclusion, we investigated the effects of collective VSC in the Raman scattering of organic polymers. Using a series of increasingly complex models, we demonstrated that the main effect of VSC is a redistribution of the total Raman cross section, with the Stokes lines for (multiple) vibrational excitation splitting into multiplets corresponding to (multiple) excitation of the lower and upper polaritons. The total cross section (integrated over the emission frequency) is approximately conserved. Using a simple analytical argument, we showed that this is true as long as the ultrastrong-coupling regime is not reached. Once the Rabi splitting becomes comparable to the transition frequency and ultrastrong coupling is achieved, the induced change in the ground state does lead to an increase of the total integrated Raman cross section, with an enhancement by less than a factor of two for realistic values. We additionally found that the Stokes lines in the strongly coupled Raman spectrum are located at the same energies (and thus posses the same Rabi splitting) as in the transmission spectrum.

In contrast, a recent experiment~\cite{Shalabney2015a} found a large enhancement of the Raman signal by two to three orders of magnitude under VSC, as well as an increase in the Rabi splitting between lower and upper polariton by more than a factor of two in the Raman spectrum compared to the transmission spectrum. We thus finish by discussing additional effects that could affect Raman scattering under VSC, and examine whether they can explain the discrepancy between theory and experiment.

First, in our models we only included a singly cavity mode, while a planar cavity supports a continuum of photonic modes. However, the argument based on \autoref{eq:totraman_hamilt} above does not depend on the number of cavity modes or molecules in the system. We have additionally confirmed this by explicitly including multiple cavity modes within the three-level model (not shown). We also neglected the rotational degrees of freedom of the molecules. The counterrotating coupling terms responsible for ultrastrong-coupling effects could lead to orientation of the molecules along the cavity-field polarization axis (if such an axis is well-defined). However, it has been shown recently~\cite{Cwik2015} that molecular orientation under strong coupling depends only on the single-molecule coupling strength without collective enhancement, such that this effect is negligible under realistic experimental conditions. Additionally, we did not consider that the molecular states could possess permanent dipole moments, which enable dipole transitions that do not change the state. We have checked explicitly that including these transitions also does not lead to an increase of the integrated Raman cross section under strong coupling.

One remaining possibility for explaining the increased Raman yield observed in the experiments within linear response is that an (unknown) VSC-induced process could lead to a modification of the bare-molecule dipole transition strengths. This would require an increase by a factor of about $\sqrt[4]{1000} \approx 6$ for each of the dipole moments, $\mu_{ge}$ and $\mu_{ev}$. This change is not contained within the state modifications induced by ultrastrong-coupling that are fully incorporated in our modeling. Nevertheless, the increase in the dipole strengths would not provide an explanation for the increased Rabi splitting observed in Raman vs.\ transmission spectra.

Finally, we have up to now neglected nonlinear effects, and only calculated the linear response of Raman scattering. Typically, Raman cross sections are quite small and nonlinear effects are thus negligible under weak coupling. However, under strong coupling, the number of populated final states reached by Raman scattering is drastically reduced, from (within the first Stokes line) one per molecule to just two extended polaritons. If the effective polariton excitation rate becomes faster than its lifetime, this could lead to an accumulation of polaritons and, subsequently, bosonic enhancement of the Raman scattering. These nonlinear interactions could also induce polariton energy shifts, such that nonlinear behavior could possibly explain both the experimentally observed enhancement as well as energy shift under Raman scattering. While a more detailed treatment is outside of the scope of this paper, it should be noted that order-of-magnitude estimates indicate that in the experiments~\cite{Shalabney2015a} excitations are created significantly slower than the polariton decay rate. Therefore, nonlinear behavior would only be expected if there is an additional enhancement factor in the system independent of strong coupling (such as, e.g., the presence of local field enhancement at hot spots if the mirror surfaces have rough structure). This highlights the necessity for further theoretical and experimental exploration of nonlinear effects in Raman scattering processes under vibrational strong coupling.

\acknowledgments
We thank A.~Shalabney and T.~Ebbesen for helpful discussions. This work has been funded by the European Research Council (ERC-2011-AdG proposal No. 290981), by the European Union Seventh Framework Programme under Grant Agreement FP7-PEOPLE-2013-CIG-618229, and the Spanish MINECO under Contract No.~MAT2014-53432-C5-5-R.

\bibliography{references}



\end{document}